\documentclass{interact}
\usepackage{epstopdf}
\usepackage[caption=false]{subfig}
\usepackage[numbers,sort&compress]{natbib}
\usepackage{pdfpages}

\bibpunct[, ]{[}{]}{,}{n}{,}{,}

\makeatletter
\def\NAT@def@citea{\def\@citea{\NAT@separator}}
\makeatother

\begin{document}


\title{Skyrmions in magnetic materials}

\author{
\name{Tom Lancaster\textsuperscript{a}\thanks{CONTACT Tom Lancaster. Email: tom.lancaster@durham.ac.uk}}
\affil{\textsuperscript{a}Department of Physics, Durham University,
  South Road, Durham, DH1 3LE, UK}
}

\maketitle

\begin{abstract}
  Skyrmions are vortex-like textures of magnetic moments found
  in some magnetically ordered materials. Here we describe the
  origin of the skyrmion and discuss the experimental realization of skyrmions in
  magnetic materials, the
  dynamic properties of the skyrmion and the potentially useful
  interaction of skyrmions with
  electrons. Our discussion is based on the physics of fields and
  allows us to touch on notions from field theory and topology that
  illustrate how skyrmions can be understood as part of a family of
  defects in an ordered field, and hint at other topological  objects that
  await our discovery. 
\end{abstract}


\section{Introduction}

If you throw a stone into a still pond then the wave you set up on the surface of
the water dies away, both at a distance
far from the initial disturbance and after a relatively short interval
of time. However, some waves in Nature are not like
this.  They are, instead, localised in space and trapped in existence for long periods, their
removal costing a great deal of energy. There are many examples of these
long-lived, wave-like {\it lumps} \cite{coleman} and this article concerns a family
of them known as {\it skyrmions}.

Waves and lumps are excitations in fields \cite{lancaster}. A field $\phi(x)$ is a machine that inputs a
point $x$ in spacetime 
and outputs the amplitude $\phi$ of the field at that point. From the point of view
of the theory of fields, every particle in the Universe is an excitation in a
quantum mechanical field. Each electron is an excitation in a electron
field, each quark an excitation in a quark field, and so on.
In the 1970s many high-energy physicists came to regard
the lumps that occur in particle fields as
sharing a similar status to the fundamental particles. At
the same time, condensed matter physicists started treating lumps as
an inevitable feature of ordered states of matter,  representing a
defect or imperfection in an ordered state. 
\begin{figure}
  \begin{center}
    \includegraphics[width=6cm]{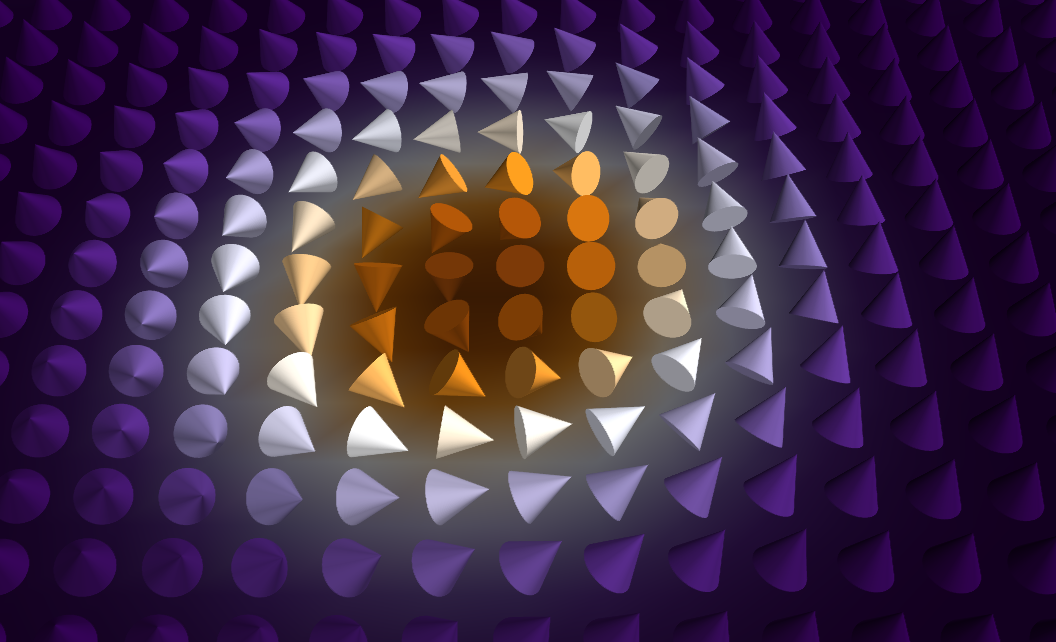}
\caption{Top: an example of a skyrmion. Bottom: the phase diagram of
  MnSi, with a small region of applied field and temperature, known as
  the $A$-phase, where a
  skyrmion lattice is observed. \label{fig:skyrm1}}
\end{center}
\end{figure}
Skyrmions are one sort of lump that
can occur in a
field whose amplitude resembles a directed arrow at each point in
space.
At the outset we will characterise lumps as
extended shapes in fields, held together by their internal interactions.
An example of a skyrmion is shown in
Figure~\ref{fig:skyrm1}, where the cones show the direction of the field
amplitude at each point. Far from the skyrmion's centre the arrows all
point in the same direction (upwards in the figure). The skyrmion itself resembles a
vortex-like shape, with the field at the centre pointing
downwards. 

A key feature of the fields that we use to construct lumps like the
skyrmion is that they must vary smoothly in space. Changing a field as a function of
spatial coordinates results in an
energy cost. However, the fields are also subject to potential-energy
costs owing to their very presence in space and time and also to their
interactions with themselves \cite{lancaster}. In magnets, these interactions can be
traced back to the properties of atoms in solids and can be understood using
arguments based on symmetry \cite{yosida}.
As usual in physics, the accountancy of
minimising the total energy is the arbiter of whether a configuration
such as a skyrmion can be
realised or not. In fact, the crucial details of the energetics in
this context can be described in terms of the overall shape of
the field in space, not just on the precise details of the field configuration. The general study of such shapes in mathematics is known
as topology,
and so lumps like the skyrmion have been given the more respectable name of {\it topological
  objects},  {\it topological defects} or {\it topological solitons} \cite{manton}. 
The topologist doesn't care about the exact distances and angles between
points on the surface of an object (the study of these is known as geometry). In fact, the topologist
imagines all objects as made from a mouldable clay that can be
stretched and shrunk, subject to the rules that holes cannot be punched
in the clay, and any existing holes cannot be healed up. Shapes that can be
transformed into each other following the rules of this game are
said to be homeomorphic to each other. There is a homeomorphism between a
coffee cup and a ring donut (that is, a torus) because in preserving the loop that forms the
cup's handle, the cup can be moulded into a donut
shape. Topologically then, the coffee cup and donut are identical \cite{lancaster}. 

In order to discover the skyrmion and the topological objects to which
it is related,  we will look at several systems distinguished by their
{\it dimensionality} $d$. Specifically, we will discuss the
one-dimensional ($d=1$) line, the two-dimensional ($d=2$) plane and the
three-dimensional ($d=3$)
space of everyday existence.
Low dimensionality (that is $d=2$ and $d=1$ behaviour) can be achieved
experimentally
in various materials in which the structure of the lattice confines the dominant interactions into one or two dimensions.
If, for example, atoms interact principally along a line
with much weaker interactions in the other two dimensions
then, within a certain realm of applicability, the solid is effectively one-dimensional.

This might all seem rather abstract, but it is the ability to realise 
skyrmions experimentally that has led to a great explosion of interest in them in
the last decade \cite{nagaosa,han, finocchio-review,sitte-review}.
Although skyrmions turned out not to be useful models of elementary particles, they have since the late 2000s been observed, produced and studied experimentally via their occurrence as excitations in the microscopic fields of magnetic materials.
Their topological 
properties turn out to endow them with an unusual set of electromagnetic
interactions. It has been argued that these properties
are potentially useful, and the control of skyrmions could be exploited
in future electronic devices. We will examine the possibility of
control of skyrmions at the end of the paper. For now,
we turn to a description of what a skyrmion is, and how they come to
exist. We begin with a little history. 

\section{A brief history of the magnetic skyrmion}

In 1961, the nuclear physicist Tony Skyrme proposed that 
topological objects could serve as a model of the
nucleons that constitute nuclear matter \cite{skyrme, manton}.
Skyrme discussed these excitations using the (jokingly named) sine-Gordon model from field
theory: a model that had originally been formulated in the completely different context of
geometry, where it is related to the description of a space with negative
curvature \cite{coleman}.
In Skyrme's model, nucleons are taken to be topological lump-like
excitations in a pion field (i.e.\ a field whose particle-like excitations
are pions).
With the emergence of the quark 
model in 1964, from independent work by Murray Gell-Mann and George 
Zweig, it was realised that neither nucleons nor pions were elementary 
particles, with both having quark constituents. Although skyrmions 
turned out not to model elementary particles they are still being 
studied today as an approximate description of atomic nuclei, with the 
original Skyrme model being developed to improve the match to low-energy 
properties of nuclei.

However, as an excitation in a field, the skyrmion can also be realised in a number
of different contexts. 
The methods of 
field theory that had been important in the development of high energy physics
were also applied across solid state (and later condensed
matter) physics, following the work of Lev Landau and collaborators in
the 1940s-60s \cite{landau}. Since then, skyrmions have regularly been suggested as possible
excitations of a number of condensed matter systems such as $^{3}$He
films, quantum Hall fluids,
superconductors and liquid crystals \cite{han}.

In magnetic systems several interactions that can stabilise skyrmions have been
identified and lead to the prediction of skyrmions with sizes ranging
across a number of different length
scales.  These include (i) the
dipolar interaction in thin magnetic films, which leads to magnetic bubbles: textures of
characteristic size 100~nm--1~$\mu$m which can be related to
skyrmions. There are also (ii) frustrated and (iii) multiple-spin
exchange interactions which can result in skyrmions whose size is
comparable to a lattice spacing.
The main subject of this article is skyrmions' proposed existence and
subsequent observation in magnetic materials with chiral crystal
symmetry, where skyrmions have a characteristic size of $5-100$~nm.
As we shall see, in magnets, skyrmions are related to  helically twisted structures of magnetic
moments. These helical arrangements can be understood
using a model of the energetics of the magnet (the Bak-Jensen model, formulated in
1980 \cite{bak}) that
invokes
the Dzyaloshinskii-Moriya interaction, which is a hybrid effect that combines
magnetic exchange and spin orbit interactions, to explain the origin of the
twisting magnetic structure \cite{dzyaloshinskii,yosida}. 
The key link to skyrmions was made by A.N.\ Bogdanov and collaborators
in the late 1980s, who
predicted the formation of a regularly repeating lattice of skyrmions
in noncentrosymmetric\footnote{A noncentrosymmetric crystal is one where
  points at coordinates $(x,y,z)$ are not equivalent to points at
  coordinates $(-x,-y,-z)$. This is the case for a crystal that is
  chiral, by which we mean the arrangement of atoms has a particular
  handedness.} (or chiral)
magnetic materials \cite{bogdanov} resembling the vortex lattice in
type II superconductors. 

The experimental observation of magnetic skyrmions was relatively recent.
In the noncentrosymmetric magnet MnSi there is a small region of
applied magnetic field and temperature that corresponds to a region of
the phase diagram known as the anomalous phase or
$A$-phase. In 2009,
small angle neutron scattering (SANS) measurements allowed the
identification of this phase as hosting a lattice formed from skyrmions \cite{muhlbauer}. This was closely
followed by observations of a skyrmion lattice in (Fe,Co)Si using microscopy
techniques \cite{yu}. 
Since then, magnetic skyrmions
have been observed in a number of 
magnetic materials (examples include FeGe, Cu$_{2}$OSeO$_{3}$ and
GaV$_{4}$S$_{8}$) \cite{nagaosa, sitte-review}
and contexts (thin magnetic layers, near
interfaces in heterostructured magnets, in patterned
nanostuctures) \cite{finocchio-review,sitte-review}. There have also been
observations of more complicated magnetic structures related to
skyrmions and recent results suggesting skyrmion phases might exist
in larger classes of material where they are stabilised by frustrated
spin interactions \cite{frustrated}. 

\section{Topological defects in ordered magnets}

In this section we examine the physics of skyrmions, with the goal of
showing
what a skyrmion is and how it arises. We will do this by building a small family
of topological objects that can exist as so-called {\it defects} in
ordered magnets \cite{blundell,chaikin,lancaster}. 

\begin{figure}
  \begin{center}
    \includegraphics[width=7cm]{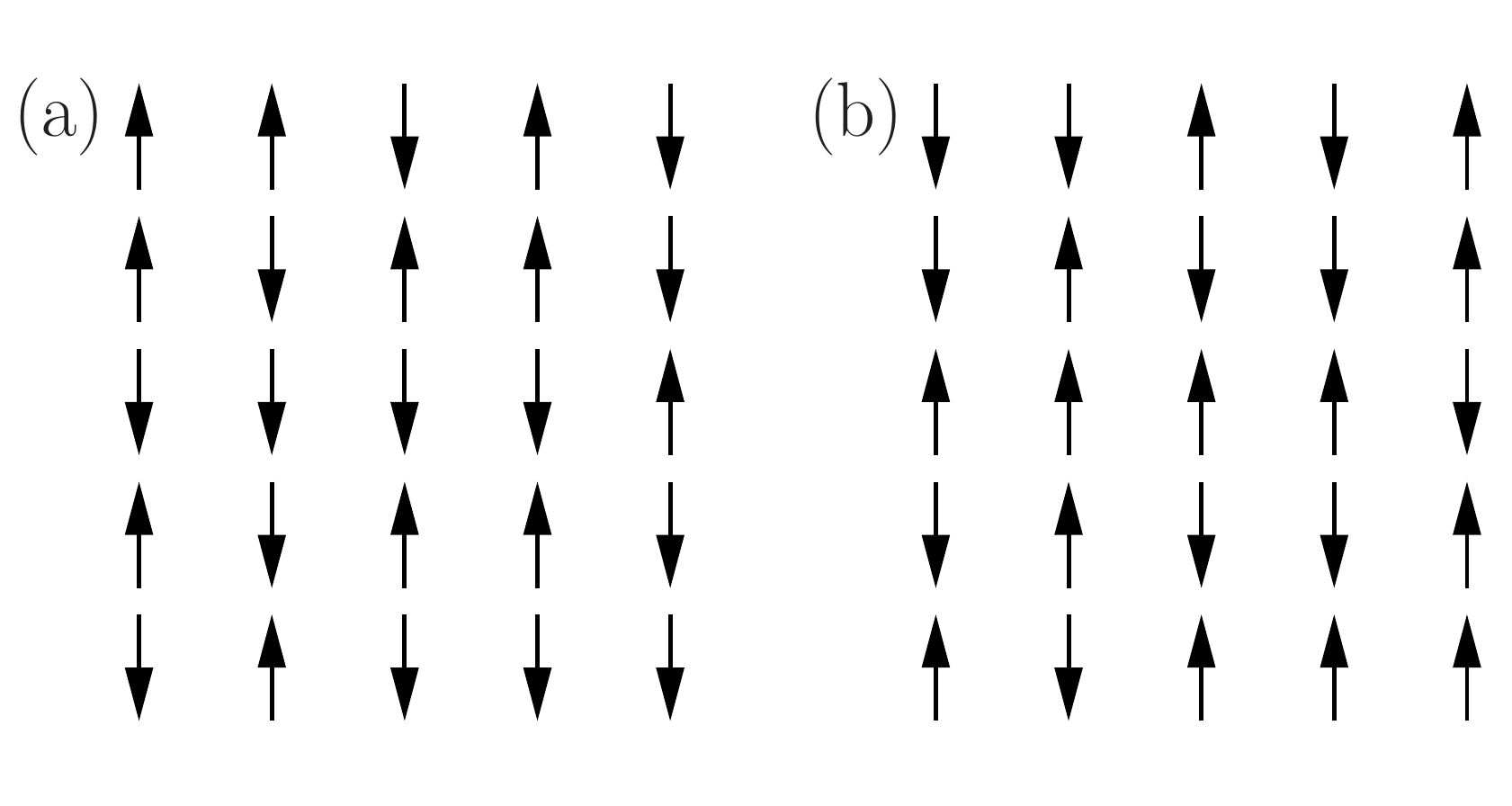}\\
    \includegraphics[width=7cm]{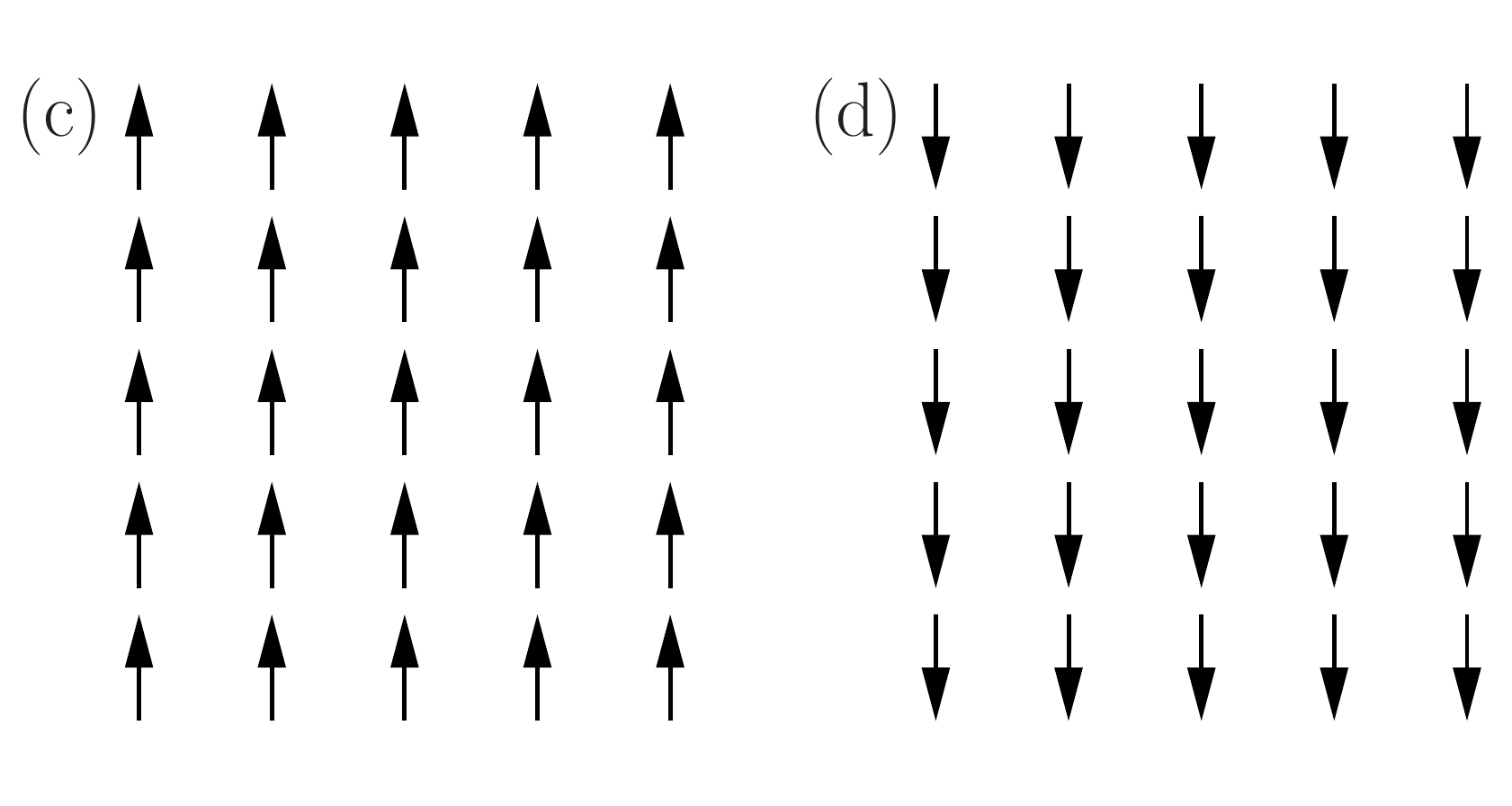}
\caption{(a) A disordered magnetic state. This configuration has magnetisation
  $M=0$ and so does (b), which is produced by turning all of the
  spins through 180$^{\circ}$. (c) An ordered magnetic phase
  with $M=M_{0}$. Turning the spins through 180$^{\circ}$ results in
  (d) which has $M=-M_{0}.$\label{fig:magnet1}}
\end{center}
\end{figure}

A simple cartoon picture of a disordered magnet is shown in
Figure~\ref{fig:magnet1}(a). The arrows represent spins (or magnetic moments) on atoms,
arranged on a lattice.
The system can be understood via its symmetry:
the magnetisation $M$ (that is, the average magnetic moment) is zero, since as many arrows
point up as down. If we turn each of the
moments through 180$^{\circ}$ then we obtain the situation shown in  Figure~\ref{fig:magnet1}(b), which also has
$M=0$. We therefore cannot tell from the magnetisation that we have transformed
the system. This inability to tell that a change has been made is known as a symmetry. 
In contrast to magnetic disorder, an ordered magnet is shown in
Figure~\ref{fig:magnet1}(c) and (d).
This has magnetisation $M_{0}\neq 0$ and 
has lost its previous symmetry in that we can now tell from the magnetisation
if a rotation is made to the arrows. We can see this by again turning the arrows
through 180$^{\circ}$, which reverses  the magnetisation $M_{0}\rightarrow
-M_{0}$. We say that the symmetry has been broken on magnetic ordering.
In Nature, this sort of ordering is observed to
takes place via a magnetic phase transition at a temperature $T_{\mathrm{c}}$.

\begin{figure}
  \begin{center}
\includegraphics[width=7cm]{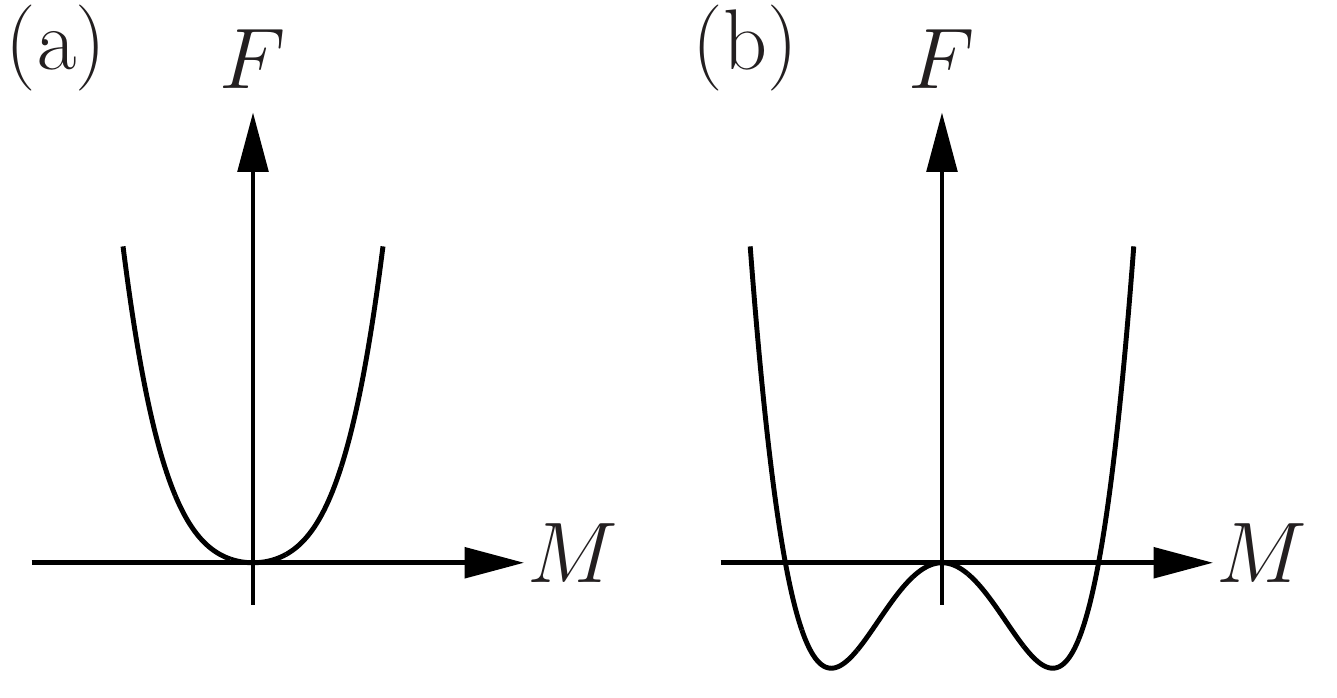}\\
\includegraphics[width=7cm]{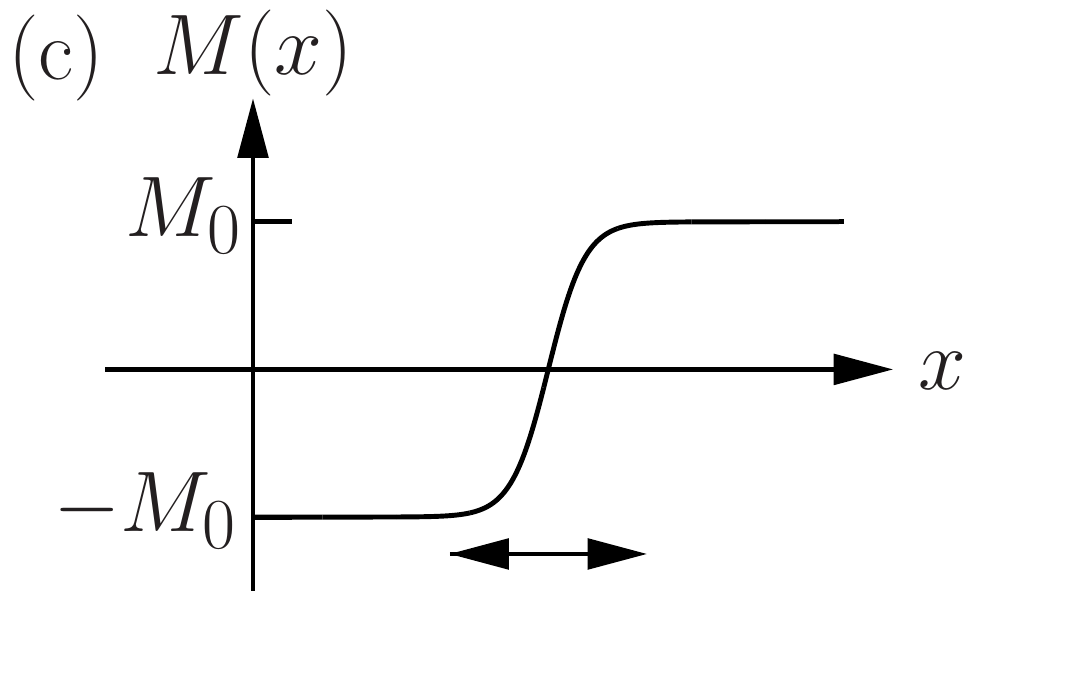}
\caption{The Landau free energy for (a) $T>T_{\mathrm{c}}$ and
  (b) $T<T_{\mathrm{c}}$. (c) The kink, or domain wall, configuration, where
  the left-hand side of the system sits in the minimum at $-M_{0}$
  and the right at $M=+M_{0}$. The spatial extent of the kink is
  shown by the double-headed arrow.\label{fig:landau1}}
\end{center}
\end{figure}

A simple mathematical description of a magnetically ordered state is
provided by the Landau model \cite{blundell,chaikin,landau,lancaster},
which describes the magnet via a free energy
\begin{equation}
F(T,M) = F_{0} + a_{0}\left(T-T_{\mathrm{c}}\right)M^{2} + \lambda
M^{4}, \label{eq:lfe}
\end{equation}
where $T$ is temperature and $a_{0}$ and $\lambda$ are positive constants. The
inclusion of only even powers of $M$ in this function ensures that the function $F$ is
symmetrical with respect to the reversal of the magnetisation.
This free energy is plotted for $T>T_{\mathrm{c}}$ and
$T<T_{\mathrm{c}}$ in Figure~\ref{fig:landau1}, where we see that the
number and position of its minima change significantly with $T$,
depending on whether we are above or below the critical temperature $T_{\mathrm{c}}$. 
This is so important because the equilibrium state of the system is found by minimising this free
energy function. From Figure~\ref{fig:landau1}(a) we can see that it
is minimised by $M=0$ for $T>T_{\mathrm{c}}$,
which corresponds to a magnetically disordered state.  For
$T<T_{\mathrm{c}}$ the free energy has two minima, found at $M=\pm
M_{0}$,
where $M_{0}\neq 0$, 
corresponding to the system aligning spins up  ($M=+M_{0}$) or down
($M=-M_{0}$). It is worth noting for later that the term $\lambda
M^{4}$ is essential here for forming the two minima for
$T<T_{\mathrm{c}}$ and, as a result, for stabilising magnetic
order. For this reason it is sometimes called the stability term. 

This picture of an ordered system was originally formulated by Lev
Landau \cite{landau} and its use in modern Condensed Matter Physics was further popularised in the
1970s and 80s by Philip Anderson \cite{anderson}, who teaches us to look for a number of key
features when a symmetry is broken. The important one for our
purposes is the notion of a {\it defect}.
In the ordered state described above, the average magnetisation field $M(x)$ is a constant, independent of position $x$. (This assumes we
average the magnetisation over blocks of the material such that the
graininess of individual atoms is unresolvable.) Constant $M(x)$ corresponds
to the whole system sitting in one of the potential minima in
Figure~\ref{fig:landau1}(b). We can, however, ask an interesting question: what if
half of the magnet (the left half, say) fell into the minimum on the
left (with $M=-M_{0}$) while the right half fell into the minimum on
the right (with $M=M_{0}$)?   We would then have the state of affairs shown in 
Figure~\ref{fig:landau1}(c), where the field changes its value as a
function of position $x$. The magnetisation field must vary smoothly
and so it passes from spin down ($-M_{0}$) on the left to spin up
($M_{0}$) on the right, making the characteristic shape
shown, which is known as a {\it kink} or {\it domain wall}. 
The kink is very stable since to remove it would cost us a semi-infinite
amount of energy. This is because it would require our lifting each moment (on the left, say) over the hump
in the
potential shown in Figure~\ref{fig:landau1}(b), in order to make the
field uniform. The kink does, however, lie a finite energy above the
ground state, which is a uniform field configuration. This is because it costs
energy to vary the field as a function of spatial coordinate (a
phenomenon known as {\it rigidity}, which is another property that
Anderson tells us to expect at all symmetry-breaking
phase transitions). The result of the energetic considerations
is a kink of finite size, as shown in Figure~\ref{fig:landau1}(c).

The energetics of a defect can be evaluated in an upgrade of our free energy that
will be useful to us later.
We use a magnetisation field
$\boldsymbol{\phi}(\boldsymbol{x})$, which is a measure of the magnetisation evaluated at
a point. (It is conventional here to switch notation to
describe the spatially variation in terms of a field $\boldsymbol{\phi}(\boldsymbol{x})$ rather
than in terms of the magnetisation $M$ we had before.) 
The free energy is $F = \int\mathrm{d}^{d}x\,f(\boldsymbol{\phi})$,
where $d$ is the dimensionality of the space and the free energy
density $f(\boldsymbol{\phi})$ is given by
\begin{equation}
f(\boldsymbol{\phi}) = c(\boldsymbol{\nabla}\boldsymbol{\phi})\cdot (\boldsymbol{\nabla}\boldsymbol{\phi}) +
a_{0}(T-T_{\mathrm{c}}) \left(\boldsymbol{\phi}\cdot\boldsymbol{\phi}\right)
+ \lambda \left(\boldsymbol{\phi}\cdot\boldsymbol{\phi}\right)^{2},
\label{eq:free_energy1}
\end{equation}
where $c$ is another positive constant. Notice how the first term tells
us that it costs energy for the field to vary in space. The other
terms resemble the ones we considered before in equation~\ref{eq:lfe}. 

The example discussed above of the domain wall is essentially a one-dimensional
one (with the field $\phi(x)$ varying in its size as we follow it
along $x$ from left to
right).\footnote{There is a slight disconnect here between the discrete
  world of spins and the continuous one of fields. Generally we
  imagine spins as each having a fixed magnitude or length, which rotate or
  flip, but don't change their length in space. Recall, however, that
  we average over many spins to get the field
  $\boldsymbol{\phi}(\boldsymbol{x})$ and so $|\boldsymbol{\phi}(\boldsymbol{x})|$ can take any
  value, reflecting the number of up and down spins in the volume over
which we average.}
In different numbers of dimensions we can produce a
family of excitations all related to the kink. Here it will be important to
distinguish between the dimensionality $d$ of the space (or atomic lattice in a
solid) and the dimensionality $D$ of the field $\phi$. The kink exists
for $d=1$ and $D=1$, which is to say we need only consider a 1$d$ line
of spins where each spin is 1$D$ in that it can point either up or down.

\begin{figure}
  \begin{center}
\includegraphics[width=7cm]{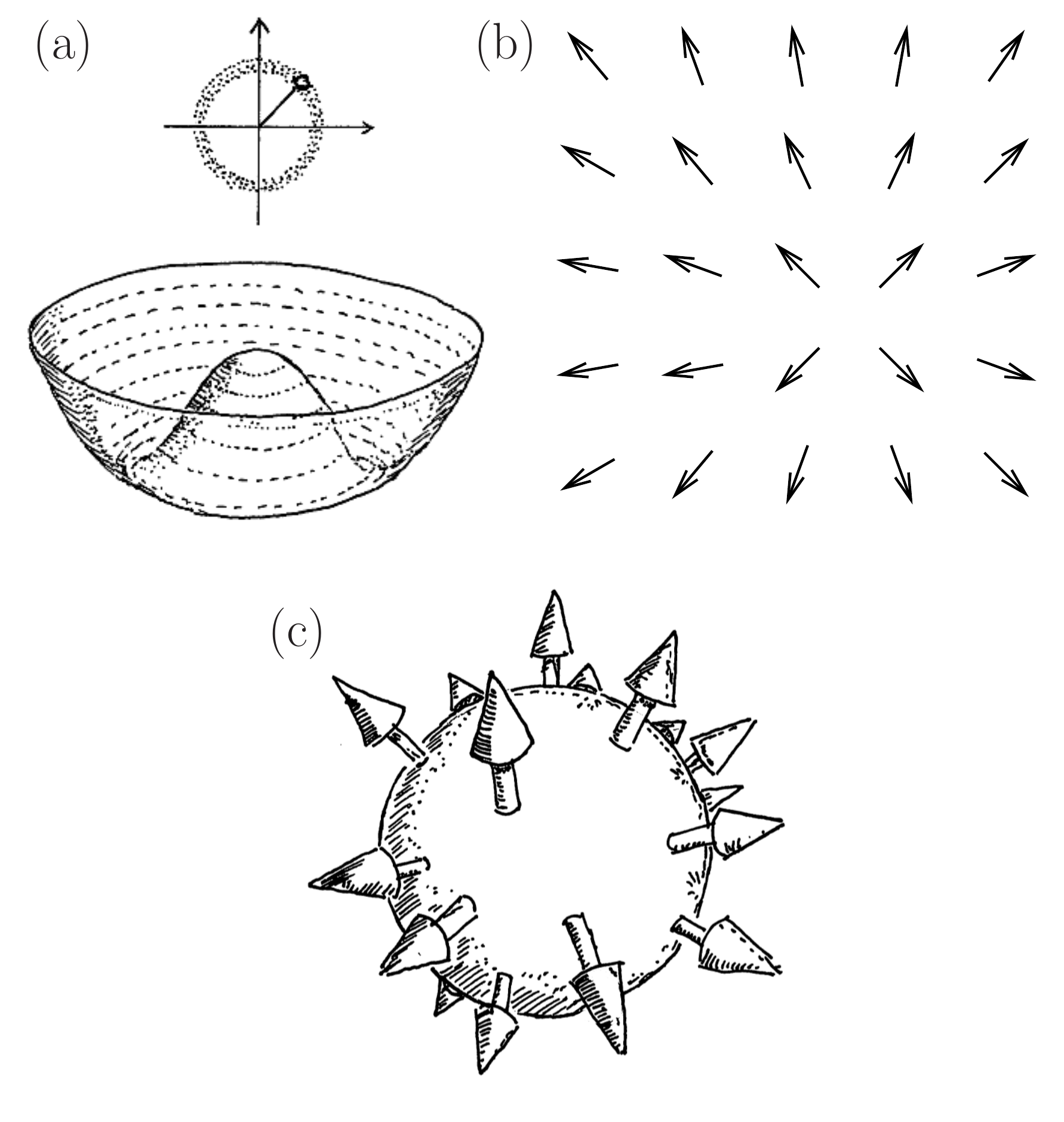}
\caption{(a) The potential energy for $d=D=2$. (b) The vortex
  excitation. (c) The hedgehog or monopole excitation for
  $d=D=3$. (Adapted from Ref~\cite{lancaster}.) \label{fig:twod}}
\end{center}
\end{figure}

Next consider the case of $d=2$ and $D=2$, that is, two-dimensional
space
and a field that can point in any direction in a
2$D$ plane. The analogue of the
double-well potential in Figure~\ref{fig:landau1}(b) resembles the bottom of a punted wine bottle or
a Mexican hat, as shown in Figure~\ref{fig:twod}(a). 
We again ask what sort of excitation can live in this potential. Like
the kink,
the answer involves the field breaking symmetry by sitting in a different
minimum of the potential in different regions of space, subject to the
all-important constraint that the field vary smoothly from place to
place. The defect in this
case is known as a {\it vortex}, an example of which is shown in
Figure~\ref{fig:twod}(b). The vortex has a core at its centre and has a
field that swirls around the core.\footnote{There is an important
  subtlety here related to the topology: the field must go to zero at the vortex core
  because the  direction of $\boldsymbol{\phi}$ is not defined at that
  point. This property of going through zero also occurs for the kink (as
  it goes from $-M_{0}$ to $M_{0}$) and will occur at the centre of the hedgehog
described below. The existence of these zeros is described
mathematically by the so-called Poincar\'{e}-Hopf theorem. } 

An important point about the vortex is that there are lots of very
similar structures we can make, for example, by globally rotating all of the arrows
by some fixed angle. In fact, from the point of view of topology
each of these excitations is equivalent. The quantity
that defines the topological properties of the vortex is its integer {\it
  winding number} $w$. This quantity counts the number of times the
arrows rotate through $2\pi$ radians as we follow a circle around the
vortex core. The diagram shows a $w=1$ vortex, since the arrows make a
complete rotation as we follow a circle around the vortex core. It is
possible to make vortices with $w=2$. In contrast to a $w=1$ object, a $w=-1$ object, known as an antivortex, doesn't have the
arrows pointing in the opposite direction, but rather has arrows that
wrap in the opposite direction as the circle is traversed around the
core.

\begin{figure}
  \begin{center}
    \includegraphics[width=8cm]{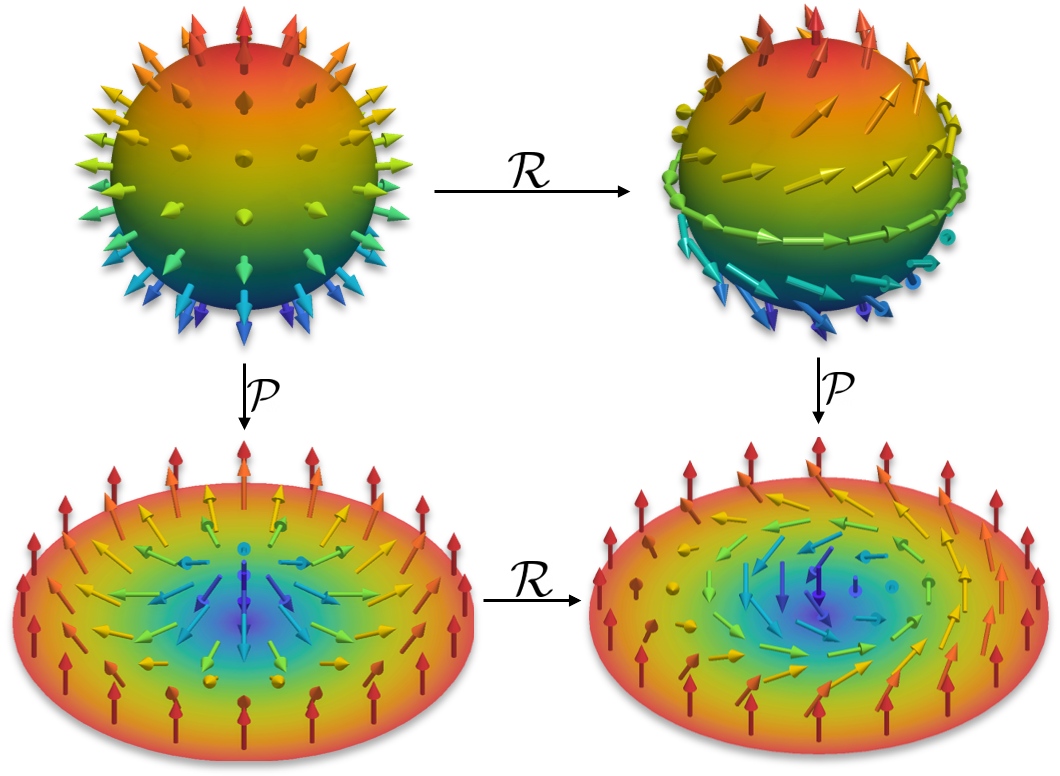}
\caption{The stereographic projection (denoted $\mathcal{P}$) squashes the hedgehog into $D=2$,
  where it becomes a skyrmion. The left-hand version is a N\'{e}el
skyrmion; the right-hand version, where the spins have been combed
over (denoted $\mathcal{R}$), is a Bloch skyrmion. (Based on a
figure from Ref.~\cite{schutte}.)\label{fig:project1}}
\end{center}
\end{figure}

In the three dimensional case of $D=3,d=3$ we have a configuration
called a hedgehog (or monopole) shown in Figure~\ref{fig:twod}(c).
Here the winding number is given by considering the 3$D$ field
$\boldsymbol{\phi}(x^{1},x^{2})$, where $x^{1}$ and $x^{2}$ are coordinates
allowing us to locate points on a closed surface (conventionally we 
choose angles $x^{1}=\theta$ and $x^{2}=\varphi$, for example), and we
evaluate the integral
\begin{equation}
w=\frac{1}{4\pi}\int\mathrm{d}x^{1}\mathrm{d}x^{2}\,
\hat{\boldsymbol{\phi}}\cdot\left(\frac{\partial \hat{\boldsymbol{\phi}}}{\partial
    x^{1}}\times \frac{\partial \hat{\boldsymbol{\phi}}}{\partial
    x^{2}}\right),
\label{eq:gauss}
\end{equation}
where $\hat{\boldsymbol{\phi}} = \boldsymbol{\phi}/|\boldsymbol{\phi}|$ is the normalised (unit)
field and
where the surface over which we integrate surrounds the core of the hedgehog. 
The integrand in this expression gives an element of the
solid angle
swept out
by the vectors $\boldsymbol{\phi}$. By comparing the integral of this
quantity with $4\pi$ we can
therefore compute how many times these vectors wrap around a sphere.
In the same way that we can globally rotate the $D=2$ arrows of the
vortex without changing $w$,  a combed hedgehog, with all of its
arrows rotated globally by the same amount, also has the same winding number as the conventional
hedgehog [see Figure~\ref{fig:project1}, top].

The vortex and hedgehog introduce a new feature compared to the domain
wall: they cost an infinite
amount of energy! This can be understood by inspection of the
vortex. It is swirly at large distances from the core, so that the fields never
becomes uniform. The first term in equation~\ref{eq:free_energy1} then
keeps costing energy causing a volume integral over the free energy density to
diverge. 
This energetic cost is a consequence of {\it Derrick's theorem} and is important in judging
whether each of
these objects can hope to exist. 
That is, if an object costs an infinite amount of
energy to create, it is not going to be realised in a system (at least
without
some other physical property being introduced)
\cite{manton,lancaster}. Specifically, Derrick investigated
static field configurations as they are scaled up and down in their spatial
size. If a field configuration is stable, then there is a point where the energy is
stationary with respect to such a scaling. If the field configuration
has no such stationary point then Derrick's theorem says it cannot
exist \cite{manton}. The instability of the vortex is a consequence of
the lack of any stationary point in its energy as its characteristic
size $L$ increases, since the energetic cost of the configuration
increases monotonically with $L$.
In addition to Derrick's theorem is also the problem of the core of the
vortex or monopole: this costs a lot of energy as there is a lot of swirliness
near the core.

So far, each of our three examples have involved $d=D$. However, this
does not have to be the case. Consider the case of 
 $D=3$, where field arrows can point in any Cartesian direction. This
 time we will constrain them to be of fixed length so, if placed at
 the centre of a sphere of fixed radius, their tips meet some
 point on the
 surface of the sphere. (The surface of the spherical ball of unit
 radius is called $S^{2}$ by
 mathematicians, so we sometimes say that the field lives in the space
 $S^{2}$). For the case above where $D=d=3$, we had the hedgehog.  We will now squash the
 $d=3$ hedgehog flat, so it can be accommodated 
 into space with $d=2$. This squashing is carried out mathematically using
 the {\it stereographic projection} as shown in
 Figure~\ref{fig:project1} \cite{schutte}.
 This is a mathematical method that
 projects
the south pole to the origin of a plane, the equator to a unit circle,
the southern hemisphere to the region inside the circle, and the
northern hemisphere to the region outside the circle, with the
northern hemisphere sent off to very large distances.

The shape that results from enacting this projection on the hedgehog
is our first example of a skyrmion. The skyrmion formed from all
arrows being radial is known as a hedgehog skyrmion or, more
conventionally, a N\'{e}el skyrmion. This reflects the fact that if
one considers a one-dimensional cut through the diameter of a skyrmion, the spins form
a N\'{e}el domain wall configuration, where the spin rotates through
$2\pi$ along the direction of the wall.  
This can be contrasted with the state of affairs resulting from the
projection of a combed  hedgehog. This is known as a Bloch skyrmion
since a corresponding one-dimensional cut through the diameter of this object gives a
Bloch-type magnetic domain wall where spins rotate perpendicular to the direction
of the wall.
(Note that these Bloch and N\'{e}el
skyrmions are topologically equivalent, with both having a winding
number $w=1$.)
The method of making cuts to reveal walls, allows us to understand another skyrmion configuration: this one
has $w=-1$ and is known as an {\it antiskyrmion} \cite{nayak}. This case is more
complicated in that cuts through the diameter of the skyrmion differ
from one another. There are two cuts through this object (at angles
$0$ and $\pi/2$) that give inequivalent
N\'{e}el walls  and two (at angles $\pi/4$ and $3\pi/4$) that give inequivalent
Bloch walls (shown in Figure~6 of Ref.~\cite{lancaster_skyrmion}).\footnote{The existence of
antiskyrmions provides one means of creating skyrmions: a
skyrmion-antiskyrmion pair could be sponteously created from the
$w=0$ ground state. Since the pair has a total winding number $w=0$ (i.e.\ the sum of
the winding numbers of a $w=1$ skyrmion and a $w=-1$ antiskyrmion)
this is allowed by topology.}
As with vortices, we can also, in principle, make skyrmions with
larger winding numbers. 
Although the winding numbers are useful invariants, it is more useful
to 
use the {\it skyrmion charge} $Q_{\mathrm{s}}$, \cite{han} which is the same as the winding number
up to a sign. The sign is supplied by the polarity of the skyrmion (i.e.\ the
difference between spin directions at the centre of the skyrmion and
far from the skyrmion). 


Mathematically \cite{nagaosa}, we can build a class of skyrmion textures in a $d=2$ plane
using a $D=3$ field with components
$\boldsymbol{\phi}(\boldsymbol{r}) =
[\cos\Phi(\varphi)\sin\Theta(r), \sin\Phi(\varphi)\sin\Theta(r),
\cos\Theta(r)]]$, where $\boldsymbol{r} =(r\cos\varphi, r\sin\varphi)$
labels the spatial coordinates in the 2-dimensional plane. Plugging
this into the expression for the winding number in equation~\ref{eq:gauss},
we find $w = \frac{1}{4\pi}\left[\cos\Theta(r)
  \right]_{r=0}^{r=\infty}\left[\Phi(\varphi)\right]_{\varphi=0}^{\varphi=2\pi}$. If
  the spins point downwards at $r=0$ and up at $r=\infty$, then the
  first square-bracketed term gives 2 and we define $2\pi m =
  \left[\Phi(\varphi)\right]_{\varphi=0}^{\varphi=2\pi}$, where $m$ is
  an integer known as the vorticity. We then have a winding number
  $w=m$ and we can define
  $\Phi(\varphi) = m \varphi + \gamma$, where $\gamma$ is a
  global phase that changes the direction of the spins with $\gamma =
  0, \pi$ corresponding to N\'{e}el skyrmions, while $\gamma = \pm \pi/2$
  corresponds to Bloch skyrmions.  

\begin{figure}
  \begin{center}
    \includegraphics[width=8cm]{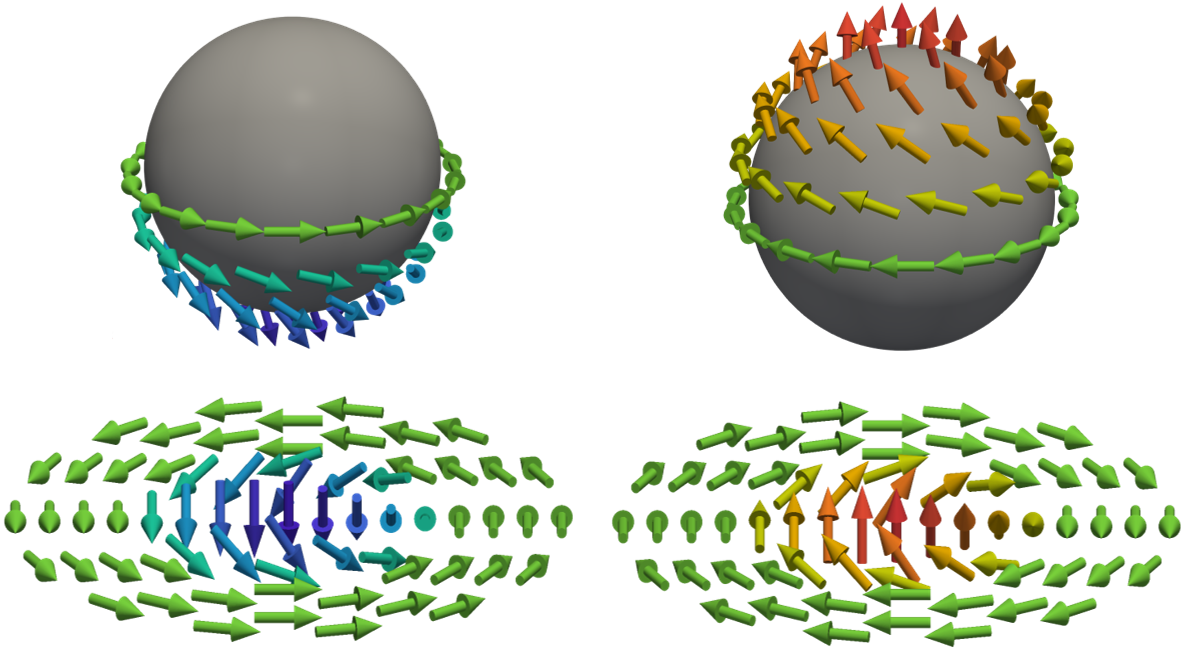}
\caption{A meron (left) results from the projection of the
  lower hemisphere of a hedgehog; an antimeron results from the
  projection of the upper half. The skyrmion can be viewed as a sum of a meron and
  antimeron. (Based on a figure from Ref~\cite{lin}).
\label{fig:merons1}}
\end{center}
\end{figure}

Another way to build a skyrmion is to consider a vortex (with $d=2$
and $D=2$) and then give
the arrows the freedom to move in a third dimension by
setting $D=3$. The spins at large distances from the core will remain
in the plane, but since the vortex core costs a great deal of energy, the
spins in the centre will develop a component in the third direction,
curling downwards (or upwards) as we move inwards towards the core. This
texture is called a {\it meron}. We can think of the structure in
terms of part of a hedgehog on a
sphere, 
whose  projection\footnote{The projection in
  Figure~\ref{fig:merons1} is not strictly the same as the ordinary
  stereographic projection in Figure~\ref{fig:project1}. 
} into $d=2$ results in the meron
[Figure~\ref{fig:merons1}] \cite{lin}. The lower
hemisphere  of the combed hedgehog gives the meron when projected into
the $d=2$ plane. From this point of view, a meron is half
a skyrmion. (It does not have an integer winding number, since the
spins don't wrap completely around the sphere.) The upper hemisphere
of the combed hedgehog is can then be used to generate an
antimeron and the skyrmion can be pictured as a meron-antimeron pair,
formed from glueing a meron and antimeron together on the sphere and
then carrying out the stereographic projection. 

\section{A brief introduction to topology and field theory}

Skyrmions are often described in terms of their topological properties
and so
here we shall very briefly touch on some of the mathematics underlying the
description of the physics in terms of topology and the consequences
for the theory of fields \cite{lancaster,altland}.\footnote{Nothing
  crucial
  will hinge on the details presented in this section, so it can be
  skipped on a first reading.}

Spaces in which topology is important are given
names: the real line is denoted $\mathbb{R}$,  the two-dimensional
plane is called $\mathbb{R}^2$ and the $n$-dimensional vector space is 
$\mathbb{R}^n$.  One can define a {\it product space} so $\mathbb{R}\times\mathbb{R}=\mathbb{R}^2$.  Similarly
$\mathbb{R}\times\mathbb{R}\times\cdots\times\mathbb{R}
=\mathbb{R}^n$.  A segment of a $\mathbb{R}$ joined end to end gives
a circle, denoted $S^1$.  As mentioned above, the sphere is called $S^2$, which is a
two-dimensional space, because by `sphere'  we mean the
surface of a ball, not its interior.  The sphere cannot be embedded
in $\mathbb{R}^2$, which is to say that you cannot fully represent a sphere on a piece of
paper without cutting it in some way.
[However $S^{2}$ can, of course, be embedded in  $\mathbb{R}^3$: there
are balls (e.g.\ tennis balls) in three-dimensional space.] \cite{lancaster}

Once we have a space, we can define a path through the
space.  If we join the ends of the path then it becomes
a loop.  Topologically, some paths can be continuously deformed into each other  and
we find that in a particular space the set of possible
loops can be divided up into a number of classes. These classes can then be
mapped onto a mathematic group (called the {\it fundamental group}, given the
symbol $\pi_1$).  For example, in $\mathbb{R}^n$ all loops are
{\it contractible} (i.e.\ they can be continuously shrunk to a point)
and so there is only one class of loop and we say that $\pi_1$ is a
{\it trivial group}
consisting of the identity element.  It is more interesting if the
space has a hole in it, so that loops can be divided into different classes,
each one
characterised by the integer number of times the loop winds round the hole.
This integer is the {\it winding number} that we met in the last
section.  As a result,
$\pi_1=\mathbb{Z}$, where $\mathbb{Z}$ denotes the set of integers.  Another very similar example is
$\pi_1(S^1)=\mathbb{Z}$ (in words: loops wrap around a circle an integer number
of times). \cite{lancaster}

In field theory, the partition function for a system is a useful
quantity that enables the average thermodynamic properties of a complicated system
to be calculated. 
The partition function can be written as a functional integral \cite{altland}
$  Z = \sum_{w}
  \int D \phi_{w}\, {\rm e}^{-S[\phi]},$
where $S$ is the action for the field theory, which is closely related
to the free energy we met in the previous section.
Often the action falls apart into two pieces
$S = S_{\mathrm{top}} + S_{0}$, where $S_{\mathrm{top}}$ is a
topological part of the action.  This topological action often takes the form $S_{\mathrm{top}} = \mathrm{i}\theta
w$, where $w$ is the winding number (again) and $\mathrm{i}\theta$ is the topological action of a field configuration with $w = 1$. In that case we have
\begin{equation}
Z = \sum_{w}{\rm e}^{-\mathrm{i}\theta w} \int D\phi_{w}\,{\rm e}^{-S_{0}[\phi]}.
\end{equation}
 This means that if we find a
theory that predicts an action
$S_{\mathrm{top}} = \frac{\mathrm{i}\theta}{4 \pi} \int\mathrm{d}x^{1}\mathrm{d}x^{2}\,
\boldsymbol{\phi}\cdot\left(\frac{\partial\boldsymbol{\phi}}{\partial x^{1}}\times \frac{\partial\boldsymbol{\phi}}{\partial x^{2}}\right),$
we expect skyrmion textures. This indeed turns out to be the case for
some theories.

In the previous sections we argued that topological objects are stable
owing to the very large energy cost that would be incurred in removing them
from existence. This is sometimes elevated to the idea of {\it topological
  protection} whereby it not possible to change the winding number of
a topological object without introducing a singularity in a field
(something forbidden by the mathematics).
It should be remembered however that this is an idealisation (often
based on infinitely large systems, or systems of reduced
dimensionality) and in
physical realisation there are often many ways of adding, removing or
altering defects, as we shall see shortly. 

This is all rather austere and involved. However, we can justify the existence of
skyrmions in magnetic materials using simpler insights from the
free energy, as we did above. We turn to this next.

\section{Chiral interactions in noncentrosymmetric magnets}

Magnetic materials are built from magnetic moments, or spins. This
discrete structure is
immediately somewhat different to our picture of continuous and smooth fields in the
previous sections. When the variation of the directions of magnetic  spins
takes place over distances that are long compared to a length $a$, the separation of
atoms, then the continuum picture of fields is likely to be a good
approximation; when this is not the case then the field picture is
expected to break down. In this latter case, changing the system in a
way that does not preserve the topology might still cost appreciable
amounts of energy, but will not be forbidden. 

At the energy
scales encountered in solids, a magnetic spin  $\boldsymbol{S}$ can be
thought of as a quantum top:
a localised arrow representing angular momentum. Spins interact in
various ways, the most 
common being the exchange interaction between two spins, which can be written
$\hat{H} = -J\boldsymbol{S}_{1}\cdot\boldsymbol{S}_{2}$,
where $J$ is an exchange constant. 
Other means of interaction are possible such as those based on 
spin orbit (SO) coupling \cite{yosida}. 
In an ion, SO coupling induces orbital moments that lead to
single-ion anisotropy: 
a contribution to the energy arising from the direction of
the electronic spin $\boldsymbol{S}$ with respect to the system's crystal axes. 
When ions interact via an exchange interaction in the presence of SO 
coupling, processes are allowed that combine these interactions. 
These interactions seem complicated at first sight, but simply rely on
considering the possible combinations, in series, of SO interactions
and exchange interactions. 
For example, starting with both
ions in their ground states, one possible process involves the SO interaction
lifting one ion out of its ground state and then the exchange
interaction returning it to the ground state. Another involves the exchange interaction
lifting an ion from its ground state and then the SO returning it. 
Taken together, the effect of these processes is to lead to an {\it
  effective interaction}
between ions that reflects the effect of both of these interactions
into an anisotropic exchange coupling
known as the {\it Dzyaloshinsky-Moriya} (DM) interaction 
\cite{dzyaloshinskii, yosida}. 
For a discrete
system of spins the Hamiltonian for the  DM interaction takes the
form 
\begin{equation}
H_{\mathrm{DM}} = \boldsymbol{D}\cdot\left( \boldsymbol{S}_{1}\times\boldsymbol{S}_{2}\right),
\end{equation}
where the so-called DM vector $\boldsymbol{D}$ depends on the details of the induced orbital
moments. 

In many solids, the DM vector $\boldsymbol{D}$ will be non-zero, although
it is strongly constrained
via Neumann's principle (that is, that the Hamiltonian shares at
least the
symmetry of the underlying crystal system). As a result, if the two ions 
have a centre of inversion midway between them, such that the symmetry
operation swaps
$\boldsymbol{S}_{1}\leftrightarrow -\boldsymbol{S}_{2}$, this 
implies $H_{\mathrm{DM}} = -H_{\mathrm{DM}}$ and so $H_{\mathrm{DM}}
=0$, implying that $\boldsymbol{D}$
must vanish. The role of $\boldsymbol{D}$ in stabilising skyrmions, and
its vanishing when there is a centre
of symmetry, explains why most putative cases of skyrmions in bulk
materials have been in
noncentrosymmetric systems. It is worth remembering though that
$\boldsymbol{D}$ is often nonzero at interfaces between materials,
leading to skyrmion textures that can be stabilised in multilayers.
For a pair of ions, it is usually possible to strongly constrain the
direction of $\boldsymbol{D}$ through symmetry arguments.
When dealing with the continuum version of the interaction appropriate
for a description in terms of fields, the same
considerations apply but may be generalised through the use of a
phenomenological approach based on
Lifshitz invariants (LIs) 
\cite{landau}. These are antisymmetric functions of the form
$\left(m_{i}\frac{\partial m_{j}}{\partial x}
-
m_{j}\frac{\partial m_{i}}{\partial x}\right)$
which occur in the free energy 
of some systems. 
The DM interaction results in a macroscopic contribution to the free energy written
$ F_{\mathrm{DM}}=   g \int\boldsymbol{\phi}\cdot(\boldsymbol{
 \nabla}\times\boldsymbol{\phi})$. 
We can then upgrade the free energy from the previous sections, so we
have \cite{han,muhlbauer}
\begin{equation}
  F =\int\mathrm{d}^{d}x\left[ c (\boldsymbol{\nabla}\boldsymbol{\phi})^{2}
  + 
a (\boldsymbol{\phi}\cdot\boldsymbol{\phi})
 + g \boldsymbol{\phi}\cdot\left(\boldsymbol{
    \nabla}\times\boldsymbol{\phi}\right) - \boldsymbol{\phi}\cdot\boldsymbol{H}+\lambda (\boldsymbol{\phi}\cdot\boldsymbol{\phi})^{2}
\right],\label{eq:ldfe2}
\end{equation}
where $\boldsymbol{H}$ is an applied magnetic field which gives the additional
contribution to the energy of $-\boldsymbol{\phi}\cdot\boldsymbol{H}$ known as the Zeeman term, that simply
reflects the fact that the spins would like to point along the
direction of the field. 

In order to see the sorts of spin fields that can exist in systems
described by this free
energy, it is helpful to switch to Fourier space,
describing the fields in terms of the sum of plane waves with
wavevectors $\boldsymbol{q}$. We take the
solid to be a finite sized box of volume $\mathcal{V}$, where the
boundary conditions imply that the waves are quantised. 
We then define a (discrete) Fourier transform
$\boldsymbol{\phi}(\boldsymbol{x}) =
\frac{1}{\sqrt{\mathcal{V}}}\sum_{\boldsymbol{q}}
\boldsymbol{
  \phi}_{\boldsymbol{q}} {\rm e}^{\mathrm{i}\boldsymbol{q}\cdot\boldsymbol{x}},$
where $\boldsymbol{\phi}_{\boldsymbol{q}}^{\dagger} =
\boldsymbol{\phi}_{-\boldsymbol{q}}$. Here $\boldsymbol{q}=0$
corresponds to a constant (ferromagnetic) spin texture. A plane wave of
 spin
density would have a single nonzero $\boldsymbol{q}$. In general,
many magnetic structures observed in Nature  can be represented by smooth
fields $\boldsymbol{\phi}(\boldsymbol{x})$ built from relatively few
distinct $\boldsymbol{q}$'s (usually $\leq 4$). 
Expressed in terms of the Fourier variables, the free energy in
equation~\ref{eq:ldfe2} is written \cite{han}
\begin{eqnarray}
  F &=&  
        \sum_{\boldsymbol{q}}
\left[
  (c\boldsymbol{q}^{2}+a)(\boldsymbol{\phi}_{\boldsymbol{q}}\cdot\boldsymbol{\phi}_{\boldsymbol{-q}})
  + \mathrm{i}g
  \boldsymbol{q}\cdot\left(\boldsymbol{\phi}_{\boldsymbol{q}}\times\boldsymbol{\phi}_{-\boldsymbol{q}}\right)
        - \boldsymbol{\phi}_{\boldsymbol{q}}\cdot \boldsymbol{H}
        \right]\nonumber \\
 && +
\lambda
\sum_{\boldsymbol{q}_{1}\boldsymbol{q}_{2}\boldsymbol{q}_{3}\boldsymbol{q}_{4}}
    \left[ \left(\boldsymbol{\phi}_{\boldsymbol{q}_{1}}\cdot
    \boldsymbol{\phi}_{\boldsymbol{q}_{2}}\right)
    \left(\boldsymbol{\phi}_{\boldsymbol{q}_{3}}\cdot
      \boldsymbol{\phi}_{\boldsymbol{q}_{4}}\right)\delta^{(3)}\left(\boldsymbol{q}_{1}
+ \boldsymbol{q}_{2} + \boldsymbol{q}_{3} + \boldsymbol{q}_{4}\right)
    \right],
    \label{eq:noncolin}
\end{eqnarray}
which is our key equation and will tell us how skyrmions can come into
being. The
second (DM) term gives us a preference for non-collinear magnetic
states (i.e.\ it makes a negative contribution to the free energy for
some swirling spin structures).
In fact, in the absence of the Zeeman term,
the ground state of this model is a helical spin structure (right
handed if $g>0$) with a wavevector of magnitude $q =g/c$.

The DM interaction therefore gives non-collinear magnetic structures as
solutions, but it is not clear yet
that it gives rise to skyrmions. As we saw before, swirling textures such as
skyrmion solutions can be killed by
Derrick's theorem. For $d=2$, the positive DM term $g\int\mathrm{d}^{2}x\,
\boldsymbol{\phi}\cdot\left(\boldsymbol{\nabla}\times\boldsymbol{\phi}\right)$
scales as $L$, the characteristic length of the region over
which the field changes, which is cause for concern, since it implies
that a skyrmion should  have $L=\infty$ if the integral is
negative (which it will be if the DM term favours swirling spin
structures).
Skyrmions therefore seem to want to unwind to save energy, with Derrick's theorem telling
us that unless the energy has a stationary point with respect to
variations in $L$, then the skyrmion cannot exist. 
However the Zeeman term $-\int
  \mathrm{d}^{2}x\,\boldsymbol{\phi}\cdot\boldsymbol{H}$ in
  equation~\ref{eq:noncolin} comes to our rescue since it scales as
  $L^{2}$, making a positive contribution since the spins forming a skyrmion
  are not lying along the field direction. 
   We then have an energy  going as
   $E = -a_{1} L + a_{2} L^{2}$, with $a_{1}$ and $a_{2}$ positive
   constants. This has a stable minimum (when $L=a_{1}/2a_{2}$),
   giving a prediction, via Derrick's theorem, of
   swirling spin structures with a finite size. One moral here is that
   a magnetic field is needed to stabilise the skyrmion. 

\begin{figure}
  \begin{center}
    \includegraphics[width=6cm]{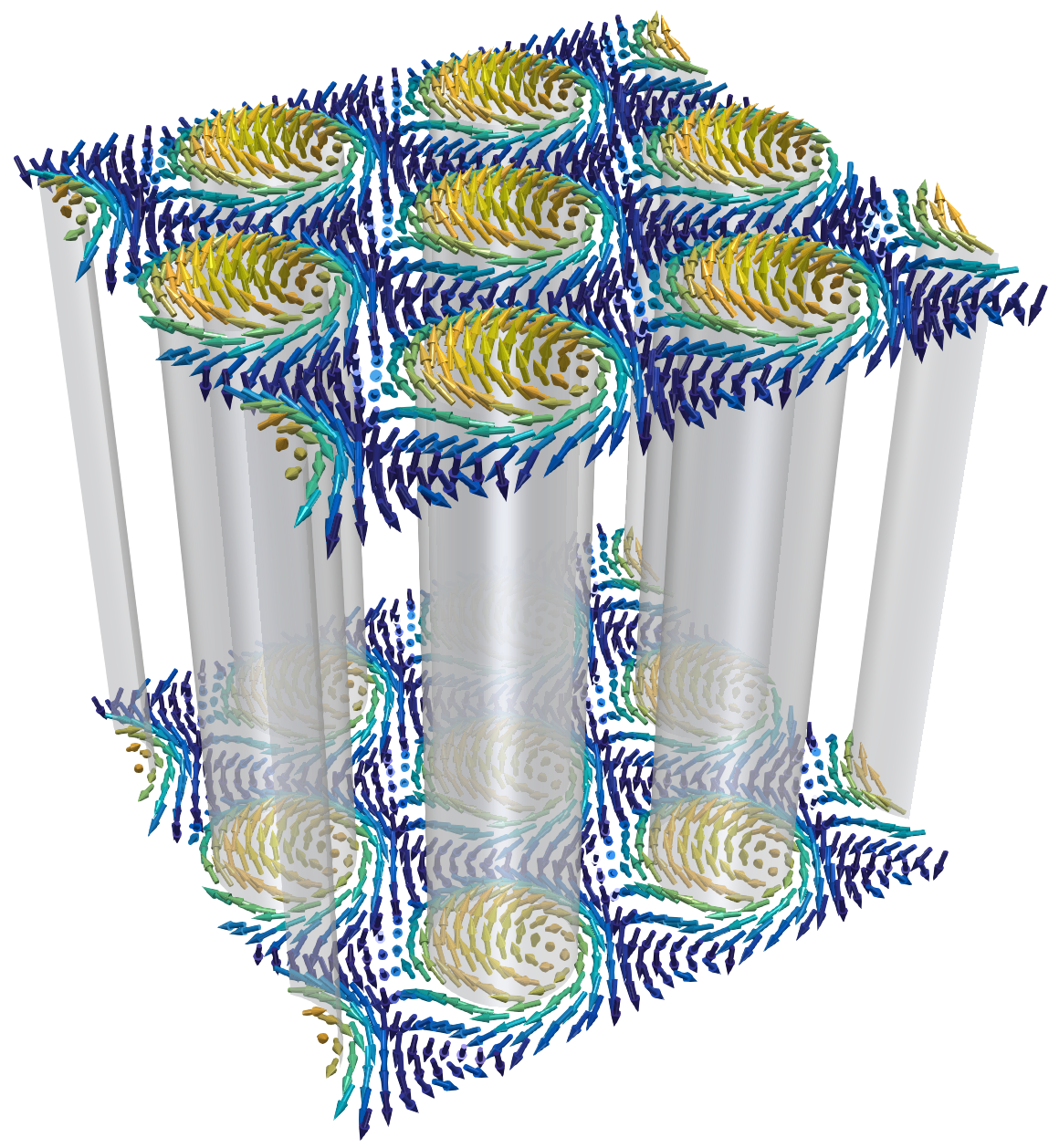}
        \caption{The skyrmion lattice with its tube-like structure in
          the third dimension.\label{fig:lattice}}
\end{center}
\end{figure}

   Two-dimensional skyrmions
  are therefore possible, at least energetically, and it turns out
that skyrmion structures that can exist stably in our model free energy can
indeed be written down.
In addition to the argument above for single skyrmions, we can
investigate the skyrmion lattice phase, since this is the one that was
observed, in MnSi for example \cite{muhlbauer}. Let's consider  the quartic term in
  equation~\ref{eq:noncolin} with prefactor $\lambda$, since this provides 
the stability of the
  broken symmetry magnetic structure in a material. 
Since the Zeeman
term is also necessary to realise skyrmions, we will try a
solution with a ferromagnetic component $\boldsymbol{q}_{1} =0$ to
reflect the Zeeman term's preference for aligned spins. With this
included, the
stability term becomes
\begin{equation}
\left[\boldsymbol{\phi}(\boldsymbol{x})\cdot\boldsymbol{\phi}(\boldsymbol{x})\right]^{2} \rightarrow
\sum_{\boldsymbol{q}_{2},\boldsymbol{q}_{3},\boldsymbol{q}_{4}}
\left(\boldsymbol{\phi}_{0}\cdot\boldsymbol{\phi}_{\boldsymbol{q}_{2}}\right)
\left(\boldsymbol{\phi}_{\boldsymbol{q}_{3}}\cdot\boldsymbol{\phi}_{\boldsymbol{q}_{4}}\right)
\delta^{(3)}(\boldsymbol{q}_{2}+ \boldsymbol{q}_{3}+
\boldsymbol{q}_{4}). 
\end{equation}
For simplicity we constrain the $\boldsymbol{q}$'s to have equal
magnitudes and end up with a contribution to the energy proportional
to 
$\delta^{(3)}(\boldsymbol{q}_{1}+\boldsymbol{q}_{2}+\boldsymbol{q}_{3})$,
implying that we seek $\boldsymbol{q}$-vectors that form an equilateral
triangle.
Just such an energetic contribution is seen elsewhere in physics in the energetics of
the liquid-solid transition \cite{chaikin,muhlbauer}. In that case there is a similar
term in the free energy that leads to a first order phase
transition between a liquid and a regular repeating crystal lattice.  
In our case this gives rise to the {\it skyrmion lattice} (SkL), which is a
triangular arrangement of skyrmions in a two-dimensional plane with
the arrangement copied in the layers of atoms in the third dimension,
such that a tube-like structure is set up in three dimensions. This is
rather like the vortex lattice in type II superconductors and is shown
in Figure~\ref{fig:lattice}.  It can be
   demonstrated via simulations \cite{muhlbauer} that thermal fluctuations make the skyrmion
   lattice phase the  equilibrium state of our model in a small pocket of
   the phase diagram close to the critical temperature $T_{\mathrm{c}}$.  

\section{Realising and detecting magnetic skyrmions}

The SkL phase was initially found experimentally in a relatively small number of materials
in a rather restricted set of
circumstances: typically a narrow region of
temperature and applied
magnetic field, usually within a few Kelvin below the magnetic
ordering temperature. 
The skyrmion phase is usually
surrounded in the temperature-applied field phase diagram by other,
non-collinear, helimagnetically ordered phases.

It seemed from the first results on bulk systems that
skyrmions were not particularly stable, owing to this very small extent
of the SkL phase in the phase diagram. However, the stability of the 
skyrmion state seems to depend essentially on the dimensionality of
the system and it was found that in 
the two-dimensional limit (corresponding to materials with a thickness of only one
skyrmion) skyrmions are observed over a far larger region of temperature
and field, with this region shrinking dramatically 
as the sample size is increased towards the three-dimensional, bulk limit.
In fact, things are not as limited as they seem in the bulk owing to another
feature of skyrmions: they can be made to exist outside of their 
equilibrium phase. As a result of fast cooling, it is possible to
temporarily trap skyrmions in existence in a  metastable state \cite{karube}.
This opens up the possibility of creating, observing and controlling
individual skyrmions outside of the confines of the often very
restricted SkL region of the phase diagram. 

A periodic structure like a lattice can be detected using diffraction
techniques and SkLs were originally measured using small angle
neutron scattering (SANS).
Although diffraction identifies periodic structures in Fourier space,
real space imaging is also possible using Transmission Electron Microscopy (TEM) 
magnetic imaging techniques (known collectively as Lorentz
microscopy). These
are based on the fact that the Lorentz force from internal magnetic
fields deflects electrons as they pass through a material, altering
the phase of the electronic wavefunction.
The resulting images that are generated can be used to measure the magnetisation of
individual skyrmions on an absolute scale.
Other experimental techniques that have been employed including x-ray
methods to measure the winding number of the skyrmion tubes, x-ray holography
and scanning transmission x-ray microscopy to image skyrmions
and muon-spin relaxation to probe the static and dynamic
field distribution of the skyrmion lattice.
Magnetometry is routinely used to measure skyrmions using AC
magnetic susceptibility. Susceptibility measures the response in the
magnetisation $M$ of a
material to a driving field $H$ given by $M=\chi H$. In AC
measurements the
idea is to apply a small, alternating component of the applied field
and measure the response $\chi'$  occurring in phase and the response
$\chi''$ occurring out of phase. It is often found that the
in-phase susceptibility $\chi'$  has a temperature- and field-dependent
magnitude that correlates closely with the skyrmion lattice phase,
while changes in $\chi''$ also marks out the skyrmion lattice phase
boundaries \cite{nagaosa}.

The first discoveries of magnetic skyrmions were made in several materials with
the noncentrosymmetric ``B20'' crystal system (with space group $P2_{1}3$). Although it was
pessimistically suggested at the time that this might be the only set of materials that
could host them, skyrmions have since
been found in bulk crystals of many other  
systems. 
As described in Section 2, the initial observation of the SkL was made in
the $A$-phase of MnSi
in 2009 using SANS, when M\"{u}hlbauer and co-workers \cite{muhlbauer} observed a 
magnetic pattern with intensity maxima that had a six-fold rotational
symmetry. 
Since then SANS has been used widely to identify the SkL in bulk
systems.
After the discovery of the SkL in MnSi,  the same group then found an
$A$-phase with similar properties in the doped semiconductor
Fe$_{0.8}$Co$_{0.2}$Si \cite{munzer}. Further evidence for 
this interpretation was provided by a real-space observation of a
two-dimensional skyrmion lattice
using Lorentz transmission electron microscopy on a thin film of
Fe$_{0.5}$Co$_{0.5}$Si \cite{yu}.
 Skyrmion
 lattices have since been found in many more materials
 \cite{sitte-review},
 including the multiferroic insulator
 Cu$_{2}$OSeO$_{3}$ \cite{seki} and  can form at room 
temperature in Co-Zn-Mn metallic alloys (which is a material with a
$\beta$-Mn structure and was the first non-B20 skyrmion-hosting
material to be identified).
There are now extensive reports of a large range
of different skyrmion textures, including N\'{e}el skyrmions
(in GaV$_{4}$S$_{8}$) \cite{kesmarki}, and antiskyrmions (in
Mn$_{1.4}$Pt$_{0.9}$Pd$_{0.1}$Sn) \cite{nayak}.

In parallel with these discoveries
are a large class of skyrmion systems where the skyrmions are
stabilised near an interface in multilayer materials. The mechanism
is again based on the DM interaction, although here it is the DM
interaction that arises at an interface between two
materials. This is realised by a thin ($<1$~nm) ferromagnetic layer of material coupled to a
material such as Pt or Ir, which has a large SO interaction. The
resulting DM interaction, along with dipolar coupling, stabilises
skyrmions (often N\'{e}el type) with characteristic size of several
hundred nanometres \cite{finocchio-review, sitte-review}. 

Furthermore, although we have presented one means of realising skyrmions in the
previous section: using the
DM interaction, we should not discount the other
proposals for ways that skyrmions could be stabilised in
magnets. 
For example, the possibility of using frustration has
attracted interest  recently with several candidate
skyrmion-hosting materials identified such as centrosymmetric
Gd$_{2}$PdSi$_{3}$ \cite{frustrated}. (Here, however,  the small size of the
skyrmions is such that we might expect the field description to break down.)

Another very important experimental technique that has been used in this
context is the Hall effect: that is, through a transverse voltage observed
when current is passed along a material in a perpendicular magnetic
field.
In skyrmion-hosting materials 
we often measure very large Hall effects. These Hall effects contain
contributions that cannot be explained by the classical Hall effect
(owing to the interaction of the conduction electrons and the external
field) nor the anomalous Hall effect (due to the interaction of
conduction electrons with the magnetisation of a ferromagnetic
conductor). 
This extra contribution is known as a {\it topological Hall
effect} and results from considering the dynamics of skyrmions and
their interactions with electrons, which is the subject to which we
now turn.\footnote{It is worth noting that there is not a one-to-one
  correspondence between observation of the topological Hall effect
  and the presence of skyrmions.}

\section{Skyrmion dynamics}

Skyrmions and skyrmion lattices are not just static structures,  but are subject to a rich
range of behaviour as a function of time. While it is often useful to
imagine the skyrmion as a rigid object whose transport involves a
change in the coodinates of its centre, it should be borne in mind
that magnetic skyrmion dynamics always involve the collective
evolution of a system's magnetic spins as a function of time. The
dynamics of magnetic spins can be understood using the {\it
  Landau-Lifshitz-Gilbert} equation of motion, which is a first-order
differential equation given (in the absence of damping) by
$\dot{\boldsymbol{n}} = -\alpha\boldsymbol{n}\times\boldsymbol{h}$,
where $\boldsymbol{n}$ gives the moment direction, $\alpha$ is the
gyromagnetic ratio, and $\boldsymbol{h}$
is an effective magnetic field that combines the externally applied
field, and internal interactions in the material \cite{landau}. (This equation of
motion for the field $\boldsymbol{n}(r,t)$ is designed to look like
that for a single moment that we briefly discuss below.)

Interestingly for our purposes, the  dynamics of magnetic systems
depend on a geometrical notion: the {\it Berry 
  phase} \cite{altland}.
The Hamiltonian of a system relies on several parameters, one example being
a magnetic field.  We can imagine taking a spin-$S$ 
particle in its ground state, with its spin aligned along the direction of an
applied magnetic field vector and then
slowly rotating the magnetic field. If we are slow enough
then, during this process, the Hamiltonian describing the system 
changes, but the spin never leaves the instantaneous ground state. (Classically we
imagine the spin staying aligned
along the field direction as we rotate the field.)
During the process of rotating the field, we move it in such a way that we rotate the tip of the field vector
around a closed loop on a sphere,
so that the field direction finishes exactly where it started. 
We might expect that
the spin wavefunction will have acquired a phase factor reflecting (i)
the time
that has elapsed during the process and (ii) the energy of the particle in
its instantaneous ground state, and this does indeed turn out to be the case.
However, it also turns out that the wavefunction acquires an additional
phase $\gamma$, that depends only on the geometry of the situation
(that is, the details of the path we chose to rotate the field around; not the speed the field was
rotated around the loop). This additional phase factor is known as a {\it Berry
phase}. For a time-dependent magnetic field of constant magnitude and a
spin $S\boldsymbol{n}$, the phase $\gamma$ is given by an integral with a
familiar integrand: the one we encountered in equation~\ref{eq:gauss}. Specifically, the phase is given by
\begin{equation}
  \gamma =
  \frac{S}{2}\int\mathrm{d}x^{1}\mathrm{d}x^{2}\,\boldsymbol{n}\cdot\left(
    \frac{\partial\boldsymbol{n}}{\partial x^{1}}
    \times
    \frac{\partial\boldsymbol{n}}{\partial x^{2}}
  \right),
  \end{equation}
where we integrate over the area enclosed on the surface of a unit sphere
by the journey of the tip of the magnetic field vector \cite{altland}. 

It is interesting to note that, when considered in terms of fields, the Berry phase $\gamma$ leads to a contribution
to the equation of motion for spins that, when considered along with the Zeeman
energy (proportional to $-\boldsymbol{S}\cdot\boldsymbol{H}$), gives rise to the 
the well-known equation of
motion  for spins $\dot{\boldsymbol{S}}
\propto\boldsymbol{S}\times\boldsymbol{H}$. In other words,
the familiar first-order equation of motion describing
the quantum analogue of the classical precession of a spin resulting
from the torque $\boldsymbol{S}\times\boldsymbol{H}$, can also be
explained as resulting from this
subtle geometrical phase \cite{altland}. Perhaps unsurprisingly for objects built
from large numbers of spins, the Berry phase will also contribute to give
us the equation of motion for skyrmions, as we now describe. 

In considering their dynamics, we describe skyrmions in terms of a skyrmion coordinate
$\boldsymbol{R}(t)=(X(t),Y(t))$, which tells us the position of a skyrmion in the $X$-$Y$
plane. We can consider the motion of spin vectors as a function of
space and time $\boldsymbol{n}(\boldsymbol{r},t)$, but if the only
source of motion is the translation of a rigid skyrmion then we can write $\boldsymbol{n}(\boldsymbol{r}-\boldsymbol{R}(t))$,
and dynamics then involve the spin configuration
changing because of displacements of the skyrmion position. 
If we take the Berry phase into account along with the coupling of
spins to local magnetic fields one can derive 
a Landau-Lifshitz Gilbert equation specifically for skyrmions, given
by \cite{han}
\begin{equation}
M_{s}\ddot{\boldsymbol{R}} = -\frac{\partial V}{\partial
  \boldsymbol{R}} + G\left(
  \hat{\boldsymbol{z}}\times\dot{\boldsymbol{R}}
\right).\label{eq:llg_skyrmion}
\end{equation}
Here $M_{\mathrm{s}}$ is an effective mass of a skyrmion, reflecting
the energetic cost of its having a velocity $\dot{\boldsymbol{R}}$ and $V$ is a
potential reflecting the energy cost of the skyrmion sitting in local
magnetic fields. The parameter $G$ is known as the gyrotropic
constant and is given by  $G= 2hSQ_{\mathrm{s}}/a^{2}$ where $S$ is the spin, $Q_{\mathrm{s}}$ is the skyrmion charge and
$a$ is the underlying lattice spacing.
In fact, the equation of motion in equation~(\ref{eq:llg_skyrmion}) looks rather like the Lorentz force
law in electromagnetism $\boldsymbol{F} = q(\boldsymbol{E} +
\boldsymbol{v}\times\boldsymbol{B})$
for the forces on an
electric charge $q$ in $E$-
and $B$-fields.
In this picture $\partial V/\partial \boldsymbol{R}$ plays the
role of the $E$-field, while $-G\hat{\boldsymbol{z}}$ becomes
the $B$-field, constrained to point along the $z$-direction
(i.e.\ out of the plane of the skyrmion).
When
the mass vanishes the two terms must cancel and we have
$G\dot{\boldsymbol{R}} = -\hat{z}\times(\partial
V/\partial\boldsymbol{R})$, which implies that massless skyrmions experience a
drifting motion, moving perpendicular to the directions of ``$E$'' and ``$B$''. 


A first example of skyrmion dynamics that makes use of these ideas is the
motion of a
single skyrmion, confined to a circular disk \cite{han}. In this case,  the
potential $V$ has the form of a restoring force, deflecting the
skyrmion from the edges of the disk. The resulting motion is described
by two gyration modes which involve the skyrmion coordinate
undergoing cyclotron-like, circular motion, such that the centre of the skyrmion
orbits the centre of the disk. The two modes represent
clockwise and anticlockwise rotations, with frequencies that differ by
a factor $G/M_{\mathrm{s}}$ and thus reflect the topological charge $Q_{\mathrm{s}}$ of
the skyrmion. 
A second example is the dynamics of the entire lattice of rigid skyrmions. In this case there are two
modes of excitation: an acoustic one and an optical one. These again represent uniform clockwise
and anticlockwise rotations, but this time they are rotations of the skyrmion lattice.
 The acoustic mode is a Goldstone mode \cite{blundell,lancaster} that reflects fact that the
 lattice of 
 skyrmions breaks translational symmetry (just like the atoms in a
 crystal lattice, whose Goldstone mode is an acoustic phonon). Although by analogy with
 phonons we might expect a linear dispersion for skyrmions, the
 acoustic mode actually has a dispersion
$\omega \propto \boldsymbol{k}^{2}$, just like the spin wave
excitations of a ferromagnet. This state of affairs is actually due
the curious, quantum mechanical feature of skyrmion dynamics: the coordinates of the
skyrmion, have a non zero commutator that obeys
$\left[ X , Y \right] = \mathrm{i}a^{2}/(4\pi S Q_{\mathrm{s}})$ \cite{han}.


So far we have only looked at the dynamics of rigid skyrmions that occur due
to the movement of the skyrmion coordinate. Another class of skyrmion
dynamics involves the modes that describe the periodic deformation of
the skyrmion's own magnetic
structure \cite{han}. These deformations can be parametrised by considering a
contour around the core of the skyrmion, connecting points with a
constant spin direction $\boldsymbol{n}$, with
radius $C(t,\phi)$, where $\phi$ is the angle in the plane and $t$ is time. For the unexcited
skyrmion this  contour will usually be a circle with radius $C_{\mathrm{s}}$,
while it will have some more
complicated behaviour as a function of time and the angle $\phi$ when
excited. The mode structure of the skyrmion can then be written
\begin{equation}
C(t, \phi) = C_{\mathrm{s}} + C_{0}(t) + C_{1}(t){\rm e}^{\mathrm{i}\phi} + C_{-1}(t)e^{-\mathrm{i}\phi}+...
\end{equation}
Here the lowest order dynamic contribution $C_{0}(t)$ is a breathing mode: the skyrmion expands and
contracts as a function of time. The next-lowest energy excitations 
reflect small displacements of the skyrmion's centre from the origin,
to a position $\boldsymbol{R} = (X,Y)$ as we had before. As might be
expected, the modes $C_{1}$ and $C_{-1}$  again reflect the cyclotron
orbits  of the centre of the skyrmion. These modes are shown in
Figure~6 of Ref.~\cite{lancaster_skyrmion}. 
An advantage of this approach of permitting the skyrmion to deform is
that it provides an expression for the skyrmion's
effective mass, which is given by
$M_{\mathrm{s}} = G^{2}R_{\mathrm{s}}^{2}/J$,
where $R_{\mathrm{s}}$ is the linear size of the skyrmion, $J$ is the
magnetic exchange constant and $G$ is the gyrotropic constant. Key here
is that the mass scales with the area $R_{\mathrm{s}}^{2}$ of the skyrmion: large skyrmions
are more massive than small ones.
Experimentally, microwave resonance experiments have been used to
excite the breathing mode and the lowest order rotating modes of the
skyrmions in the gigahertz
regime \cite{onose}. Using this technique, microwaves are absorbed
at the characteristic resonant frequencies of the magnetic
structure, allowing a spectroscopic insight into the presence of
skyrmions and their dynamics. 

\section{Spin transfer torque and emergent electrodynamics}

The large contributions skyrmions give to the Hall effect that we met earlier reflects
some notable features of their electrodynamics that we discuss here. 
{\it Hund's rule coupling} is the name we give to the quantum mechanical effect where a single electron
has its spin biased in the direction of a local moment \cite{blundell}. (It is
familiar from atomic physics, where it partially explains the spin ground state
of a magnetic ion and arises from electrostatic considerations and the
Pauli exclusion principle.)
Skyrmions interact with electrons via Hund's rule coupling, which
aligns the electronic spin along the direction of the local
magnetic moment of the skyrmion $\boldsymbol{n}$. Since electrons move
very fast (that is, much
faster than skyrmion spins are moving), the electron has enough time to
completely orient its spin in the direction of $\boldsymbol{n}$. (As
described above, this can lead to a geometric Berry phase.)
Rotating electron spins requires that the skyrmion supplies a torque,
and so classical mechanics tells us
that there must be an equal and opposite torque acting on the spins in
the skyrmion. 
This leads to an effect known as the {\it spin transfer
  torque} that allows skyrmions to be moved by electric currents \cite{han,nagaosa}.
An important example is a current flowing through a skyrmion which, in the
absence of dissipation, gives rise to a drift of the skyrmion in the
direction of the current. 

We saw above [equation~(\ref{eq:llg_skyrmion})] how the forces that cause the motion of rigid skyrmions
bear a resemblance to those that result from Lorentz's force law in
electromagnetism. The rather different case of the motion of an electron that passes
through a skyrmion also enjoys a link to electromagnetism (although it is
worth keeping these two cases separate). In the current
case, where the electron's motion is determined by Hund's rule
coupling, the electron acts just as if it is passing through a  
electromagnetic field. It is
 not simply that the field straightforwardly reflects the local
 electromagnetic forces
from each of
the magnetic ions on the electron (this would need to take Pauli
exclusion into account for example); it is {\it as if} there is a
new electromagnetic field present, determined by the topology of the skyrmion.
This has been called an {\it emergent}
electromagnetic field: all of the microscopic, quantum mechanical Hund's rule couplings act in a
complicated manner and what emerges is a rather simple effective force field
that has the mathematical structure of a classical electromagnetic
field \cite{schulz,nagaosa,han}.

In conventional  electrodynamics, the electromagnetic field is
described in terms of a gauge field \cite{lancaster}
with components (in natural units, where the speed of light is set to unity) $A^{\mu}= (V, A^{x}, A^{y}, A^{z})$,
that causes a force on electric charges moving at a velocity
$\boldsymbol{v}$ with components $q
F_{\mu\nu}v^{\mu}$, where $F_{\mu\nu}$
are the
components of 
the Faraday tensor, a $4\times 4$ matrix built from the components
$A^{\mu}$ using a four-dimensional cross product, written as \cite{lancaster}
\begin{equation}
F_{\mu\nu} =
\left(\frac{\partial A_{\nu}}{\partial x^{\mu}} -
  \frac{\partial A_{\mu}}{\partial x^{\nu}}\right).
\end{equation}
For skyrmions, 
the twists and turns of the skyrmion spin texture can be thought of
as giving rise to an emergent electromagnetic gauge field
$\boldsymbol{a}$, with components $\boldsymbol{a} = (U,
a^{x}, a^{y}, a^{z})$, that
couples to the electrons 
in exactly the manner that the real
electromagnetic gauge field $A^{\mu}$
does.  In the
skyrmion case, the electrons couple to the components $f_{\mu\nu}$ of
an emergent Faraday tensor, which is determined by the spin
configuration of the skyrmion via  \cite{han}
\begin{equation}
f_{\mu\nu} =
-\frac{\hbar}{q_{e}}\left(\frac{\partial a_{\nu}}{\partial x^{\mu}} -
  \frac{\partial a_{\mu}}{\partial x^{\nu}}\right) =
-\frac{\hbar}{2q_{e}}\boldsymbol{n}\cdot \left(\frac{\partial
  \boldsymbol{n}}{\partial x^{\mu}}
- \frac{\partial \boldsymbol{n}}{\partial x^{\nu}}\right), 
\end{equation}
where $q_{e}$ is the electronic charge.
Perhaps we should not be surprised to see the now-familiar 
integrand from equation~\ref{eq:gauss} providing the components of the effective Faraday tensor. 

This abstract picture can be made more vivid by picking out the emergent
electric $\boldsymbol{e}$ and magnetic $\boldsymbol{b}$ fields from $f_{\mu\nu}$. 
Specifically we have \cite{han}
\begin{equation}
  \begin{array}{cc}
\boldsymbol{b} =
  -\frac{\hbar}{q_{e}}\left(\boldsymbol{\nabla}\times\boldsymbol{a}\right), &
  \boldsymbol{e} = -\frac{\hbar}{q_{e}}\left(
\boldsymbol{\nabla}U - \frac{\partial\boldsymbol{a}}{\partial t}
  \right).\end{array}
\end{equation}
In words, the first expression says that the skyrmion spin texture is
the source of emergent magnetic field; the second equations says that
the moving skyrmion is the source of
electric field.

\begin{figure}
  \begin{center}
    \includegraphics[width=5cm]{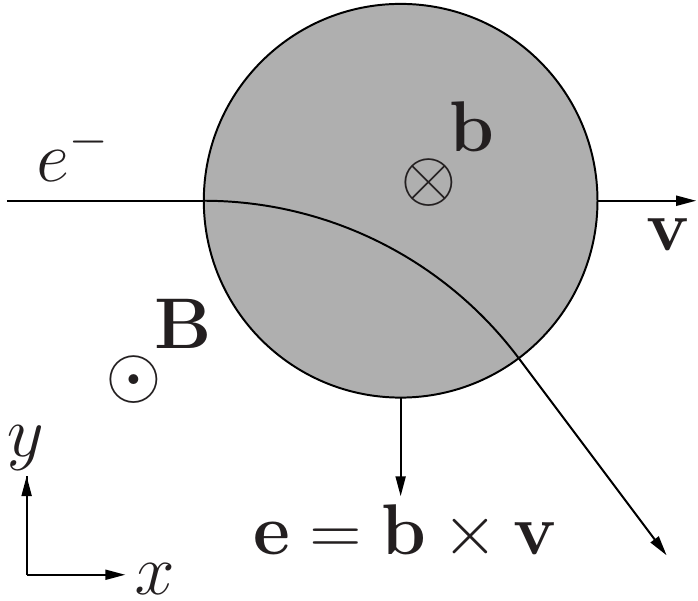}
\caption{The forces on a skyrmion when a current passes though it~\cite{han}.\label{fig:she}}
\end{center}
\end{figure}

We motivated our discussion of skyrmion dynamics by an attempt to understand
the large topological contribution to the Hall effect seen in
materials when skyrmions are present \cite{han}. 
The size of this contribution follows from the
magnetic flux $\Phi$ experienced by an electron from the presence of the
skyrmion. This is given by
$\Phi = -hQ_{s}/q_{e}$, which  turns out to be a very large field for a
typical skyrmion: typically tens of tesla. 
Following Ref~\cite{han}, let's examine an example of the motion of a
skyrmion and an electron (Figure~\ref{fig:she}).
If we apply an external $B$-field along the $z$-direction then an
electron with charge $q_{e}$
traveling along the $x$-direction with velocity $\boldsymbol{u}$ experiences a Lorentz force
$q_{e}\boldsymbol{u}\times\boldsymbol{B}$, deflecting the electron in
a direction perpendicular to $\boldsymbol{u}$ and $\boldsymbol{B}$, in
this case along $+y$.
If the applied $B$-field stabilises a skyrmion, then the 
emergent magnetic field $\boldsymbol{b}$
from a skyrmion is typically much larger than the external magnetic field $\boldsymbol{B}$, in the opposite
direction (owing to the minus sign in $\Phi = -hQ_{s}/q_{e}$),
and therefore deflects the electron exactly in the
opposite direction: along $-y$. The electron therefore experiences a large
Hall effect due to the presence of the skyrmion. 
However, the motion of electrons also
causes the skyrmion to drift in the direction of the electron flow
with velocity $\boldsymbol{v}$. Since the skyrmion is
now at motion, this causes an
emergent electric field $\boldsymbol{e} =
\boldsymbol{b}\times\boldsymbol{v}$ directed along $-y$ that (owing to
the electron's negative charge) produces
a force on the electron along $+y$. This
lessens
the amount of deflection along $-y$ compared to the case of a static skyrmion. The moving
skyrmion therefore causes a smaller Hall effect compared to a static one. 
 
\section{Conclusions}

One reason why skyrmions have received so much attention is the
possibility of using them in spintronics applications. 
Based on the discussion in the previous section,
there is a prospect of using skyrmions to provide a very efficient coupling
of an electric or spin current to a magnetic structure.
Since skyrmions carry 
a invariant skyrmion charge $Q_{\mathrm{s}}$ (or winding number $w$) it has
been suggested that this could be used to encode information. 
Compared to 
magnetic domain walls used in many magnetic memory applications,
there is experimental evidence that skyrmions could be transported at surprisingly
low energy cost. Great excitement followed from the observation that an
the spin transfer torque allowed an electrical current density of just $2.2\times 10^{6}$~Am$^{-2}$
to cause motion of the 
skyrmion lattice \cite{jonietz}. 
Although, happily, this is very small indeed compared to the $10^{11}$~Am$^{-2}$ required to move
ferromagnetic domain walls by spin transfer torque, if high
speed manipulation of the skyrmions is attempted then the current densities become
similar. 
This would seem to imply that although skyrmions might constitute
a new, low-energy avenue for magnetic data storage, logical bit-wise operations
or spintronic technologies, there is still much work to be done in learning how to
best exploit skyrmions for technology. We would, for example, prefer to
use small ($<10$~nm) skyrmions to maximise the density of data and to
realise them at room temperature and without the need for an applied
magnetic field.
As a route towards their exploitation, there has  been much discussion of the properties of skyrmions
confined in various geometries. Finite-sized systems allow different
pathways to creating and annihilating skyrmions, owing to the lowered
energy barriers at boundaries. 
A much-discussed example of a skyrmion-based data storage device is the
racetrack model \cite{sitte-review}. Here the skyrmions resemble the beads on an
abacus whose presence or absence provide the series of 1s and 0s
required to store digital information. Such a device would need to comprise  (i) a writing module based
on localised skyrmion creation;
(ii) some means of moving skyrmions; and (iii) a means of reliably detecting
them. We are some way from having this level of control, but it does
not seem unreasonable. Other ideas to utilise skyrmions include
proposals for creating logic gates and even for probabilistic computing devices \cite{finocchio-review,sitte-review}. 
With the current
excitement in the field, it is likely there will be many other
suggestions for applications as magnetic skyrmions are investigated more widely.


From the point of view of realising topological objects, perhaps
predictably, the skyrmion doesn't exhaust the possibilities of
finding complex, particle-like excitations in condensed matter systems
and beyond.
An example of a more complicated object is a $d=3$ topological excitation, known as
hopfion \cite{sutcliffe}, shown in Figure~11 of Ref.~\cite{lancaster_skyrmion}.
If we take a finite length of skyrmion tube,  each cross-section contains a $d=2$
skyrmion. If we twist one end of the tube through $2\pi$ and join it
to the other 
end,  we form a closed loop. This loop has the topology of a hopfion.
In the  $d=2$ skyrmion, one spin in the texture  points
towards each direction once. If the magnetisation points downwards at
the centre of the skyrmion then, in the hopfion we have constructed
from the skyrmion tube, the magnetisation points
downwards along a circle in space. The same is true for any other spin direction,
which also forms a loop. The defining property of the hopfion is that
each of these loops formed by following a particular spin direction
links another one only once. We say that the hopfion linking number $Q$
is one. As always, this does not exhaust the possibilities and further
multihopfions can be constructed with $Q>1$. 

We have seen how topological defects such as the skyrmion (or hopfion)
are members of a family of stable field excitations. These
can be  treated on the same sort of basis as ``fundamental'' particles
with their own set of properties and interactions. While the existence
of skyrmions in magnetic materials is now well established, their
existence as excitations in other matter and gauge fields is not so firmly founded. However, it
might well be that defects formed in the fields of the early Universe
and have resulted in detectable features that we will encounter in the
future. Cosmological speculation
aside, the idea of encoding information in domain wall defects is a very
important technological one and the potential for skyrmions to be used
low-energy-cost data devices makes it likely that they will continue
to be widely discussed in the coming years in one form or
another. Whatever one's field of interest within physics, the story of
the magnetic skyrmion suggests the usefulness and unifying
potential of notions from topology and field theory and hints that
this is an area to watch in the future.

\section*{Acknowledgements}
I am very grateful to Stephen Blundell, Martin Galpin, Matja\v{z} Gomil\v{s}ek,
Peter Hatton, Thorsten Hesjedal, Thomas Hicken, Ben Huddart,
Michael Spannowsky and Paul Sutcliffe for their many helpful comments on
earlier versions of this review.
I thank Sam Moody for all of his efforts in preparing some of the
figures and permission to use Figures~\ref{fig:project1} and \ref{fig:merons1};
Mark Vousden, Marijan Beg and Hans Fangohr (European XFEL GmbH
and University of Southampton) for permission to use
Figure~\ref{fig:skyrm1}; and Max Birch (Durham University) for
producing Figure~\ref{fig:lattice}.
I am also indebted to the
members of the UK Skyrmion Project [supported
by EPSRC (UK)] for their collaboration on this topic and to Stephen
Blundell for our  collaboration on field theory. 

\section*{Funding}
This work is supported by EPSRC (UK), grant no.\ EP/N032128/1.

{\it This is an Author's Original Manuscript of an article published by Taylor \&
Francis in Contemporary Physics on Dec 23 2019, available online:
http://www.tandfonline.com/doi/10.1080/00107514.2019.1699352.}

%

\end{document}